\documentclass[12pt,letterpaper]{JHEP3}
\usepackage{amsfonts,amssymb}
\usepackage{amsmath}
\usepackage{amsthm}
\usepackage{graphicx}
\usepackage{axodraw}
\newcommand{\pa}{\partial}
\newcommand{\tr}{{\rm tr}}
\newcommand{\comment}[1]{}

\theoremstyle{plain}
\newtheorem{proposition}{Proposition}[section]
\newtheorem{theorem}[proposition]{Theorem}

\newtheorem*{conjecture*}{Conjecture}

\newcommand{\pasl}{\pa\kern-.45em /}

\newcommand{\ksl}{k\kern-.55em /}

\DeclareFixedFont{\xiiss}{OT1}{cmss}{m}{n}{12}
\DeclareFixedFont{\ixss}{OT1}{cmss}{m}{n}{9}
\DeclareFixedFont{\cmrnine}{OT1}{cmr}{m}{n}{9}

\newcommand{\CCs}{\hbox{\ixss C\kern-.4emI}}
\newcommand{\ZZs}{\hbox{\ixss Z\kern-.4emZ}}

\newcommand{\bm}[2]{\begin{minipage}{#1}#2\end{minipage}}
\newcommand{\dpquiv}{\bm{1in}{\begin{picture}(63,31) (35,-30)
\SetWidth{0.5}
\SetColor{Black}
\ArrowLine(67,-6)(37,-6)
\ArrowLine(97,-6)(67,-6)
\Vertex(67,-6){1.43}
\Vertex(52,-28){1.43}
\Vertex(82,-28){1.43}
\ArrowLine(37,-6)(52,-28)
\ArrowLine(82,-28)(97,-6)
\ArrowLine(52,-28)(67,-6)
\SetColor{SkyBlue}
\ArrowLine(82,-28)(52,-28)
\SetColor{Black}
\ArrowLine(67,-6)(82,-28)
\SetColor{SkyBlue}
\ArrowArc(52,-17)(18,36.87,143.13)
\ArrowArc(82,-17)(18,36.87,143.13)
\SetColor{Black}
\Vertex(37,-6){1.43}
\Vertex(97,-6){1.43}
\ArrowLine(37,-6)(67,-21)
\ArrowLine(67,-21)(97,-6)
\Line(67,-21)(82,-28)
\Line(52,-28)(67,-21)
\end{picture}}}

\newcommand{\quiv}[1]{\bm{.9in}{\begin{picture}(63,31)(35,-30)\SetWidth{0.5}\SetColor{Black}
\Vertex(67,-6){1.43}
\Vertex(52,-28){1.43}
\Vertex(82,-28){1.43}
\Vertex(37,-6){1.43}
\Vertex(97,-6){1.43}
#1\end{picture}}}
\newcommand{\ba}{\SetColor{SkyBlue}
\ArrowLine(67,-6)(37,-6)
\SetColor{Black}}
\newcommand{\bb}{\SetColor{SkyBlue}
\ArrowLine(97,-6)(67,-6)
\SetColor{Black}}
\newcommand{\bc}{\SetColor{SkyBlue}
\ArrowLine(82,-28)(52,-28)
\SetColor{Black}}
\newcommand{\xl}{\ArrowLine(37,-6)(52,-28)}
\newcommand{\yl}{\ArrowLine(52,-28)(67,-6)}
\newcommand{\zl}{\ArrowLine(67,-6)(37,-6)}
\newcommand{\cl}{\ArrowLine(37,-6)(67,-21)\Line(67,-21)(82,-28)}
\newcommand{\xr}{\ArrowLine(67,-6)(82,-28)}
\newcommand{\yr}{\ArrowLine(82,-28)(97,-6)}
\newcommand{\zr}{\ArrowLine(97,-6)(67,-6)}
\newcommand{\crr}{\ArrowLine(67,-21)(97,-6)\Line(52,-28)(67,-21)}

\newcommand{\cpd}{\bm{1in}{\begin{picture}(63,31) (35,-30)
\SetWidth{0.5}
\SetColor{SkyBlue}
\ArrowLine(67,-6)(37,-6)
\ArrowLine(97,-6)(67,-6)
\SetColor{Black}
\Vertex(67,-6){1.43}
\Vertex(52,-28){1.43}
\Vertex(82,-28){1.43}
\SetColor{SkyBlue}
\ArrowLine(82,-28)(52,-28)
\SetColor{Black}
\SetColor{Black}
\Vertex(37,-6){1.43}
\Vertex(97,-6){1.43}
\ArrowLine(37,-6)(67,-21)
\ArrowLine(67,-21)(97,-6)
\Line(67,-21)(82,-28)
\Line(52,-28)(67,-21)
\end{picture}}}
\newcommand{\cpdp}{\bm{1in}{\begin{picture}(63,31) (35,-30)
\SetWidth{0.5}
\SetColor{Black}
\ArrowLine(67,-6)(37,-6)
\ArrowLine(97,-6)(67,-6)
\SetColor{Black}
\Vertex(67,-6){1.43}
\Vertex(52,-28){1.43}
\Vertex(82,-28){1.43}
\SetColor{SkyBlue}
\ArrowLine(82,-28)(52,-28)
\SetColor{Black}
\SetColor{Black}
\Vertex(37,-6){1.43}
\Vertex(97,-6){1.43}
\ArrowLine(37,-6)(67,-21)
\ArrowLine(67,-21)(97,-6)
\Line(67,-21)(82,-28)
\Line(52,-28)(67,-21)
\end{picture}}}

\newcommand{\cpcb}{\bm{1in}{\begin{picture}(63,31) (35,-30)
\SetWidth{0.5}
\SetColor{SkyBlue}
\ArrowLine(67,-6)(37,-6)
\ArrowLine(97,-6)(67,-6)
\SetColor{Black}
\Vertex(67,-6){1.43}
\Vertex(52,-28){1.43}
\Vertex(82,-28){1.43}
\ArrowLine(37,-6)(52,-28)
\SetColor{SkyBlue}
\SetColor{Black}
\SetColor{Black}
\Vertex(37,-6){1.43}
\Vertex(97,-6){1.43}
\ArrowLine(67,-21)(97,-6)
\Line(52,-28)(67,-21)
\end{picture}}}
\newcommand{\cpca}{\bm{1in}{\begin{picture}(63,31) (35,-30)
\SetWidth{0.5}
\SetColor{SkyBlue}
\ArrowLine(67,-6)(37,-6)
\ArrowLine(97,-6)(67,-6)
\SetColor{Black}
\Vertex(67,-6){1.43}
\Vertex(52,-28){1.43}
\Vertex(82,-28){1.43}
\ArrowLine(82,-28)(97,-6)
\SetColor{SkyBlue}
\SetColor{Black}
\SetColor{Black}
\Vertex(37,-6){1.43}
\Vertex(97,-6){1.43}
\ArrowLine(37,-6)(67,-21)
\Line(67,-21)(82,-28)
\end{picture}}}
\newcommand{\cpbc}{\bm{1in}{\begin{picture}(63,31) (35,-30)
\SetWidth{0.5}
\SetColor{SkyBlue}
\ArrowLine(67,-6)(37,-6)
\SetColor{Black}
\ArrowLine(97,-6)(67,-6)
\Vertex(67,-6){1.43}
\Vertex(52,-28){1.43}
\Vertex(82,-28){1.43}
\ArrowLine(37,-6)(52,-28)
\SetColor{SkyBlue}
\SetColor{Black}
\SetColor{Black}
\Vertex(37,-6){1.43}
\Vertex(97,-6){1.43}
\ArrowLine(67,-21)(97,-6)
\Line(52,-28)(67,-21)
\end{picture}}}
\newcommand{\cpbap}{\bm{1in}{\begin{picture}(63,31) (35,-30)
\SetWidth{0.5}
\SetColor{Black}
\ArrowLine(67,-6)(37,-6)
\SetColor{SkyBlue}
\ArrowLine(97,-6)(67,-6)
\SetColor{Black}
\Vertex(67,-6){1.43}
\Vertex(52,-28){1.43}
\Vertex(82,-28){1.43}
\ArrowLine(37,-6)(52,-28)
\SetColor{SkyBlue}
\SetColor{Black}
\SetColor{Black}
\Vertex(37,-6){1.43}
\Vertex(97,-6){1.43}
\ArrowLine(67,-21)(97,-6)
\Line(52,-28)(67,-21)
\end{picture}}}

\newcommand{\cpbb}{\bm{1in}{\begin{picture}(63,31) (35,-30)
\SetWidth{0.5}
\SetColor{Black}
\Vertex(67,-6){1.43}
\Vertex(52,-28){1.43}
\Vertex(82,-28){1.43}
\ArrowLine(52,-28)(67,-6)
\SetColor{SkyBlue}
\ArrowLine(82,-28)(52,-28)
\SetColor{Black}
\ArrowLine(67,-6)(82,-28)
\SetColor{SkyBlue}
\SetColor{Black}
\Vertex(37,-6){1.43}
\Vertex(97,-6){1.43}
\end{picture}}}
\newcommand{\cpba}{\bm{1in}{\begin{picture}(63,31) (35,-30)
\SetWidth{0.5}
\SetColor{SkyBlue}
\ArrowLine(97,-6)(67,-6)
\SetColor{Black}
\ArrowLine(67,-6)(37,-6)
\Vertex(67,-6){1.43}
\Vertex(52,-28){1.43}
\Vertex(82,-28){1.43}
\ArrowLine(82,-28)(97,-6)
\SetColor{SkyBlue}
\SetColor{Black}
\SetColor{Black}
\Vertex(37,-6){1.43}
\Vertex(97,-6){1.43}
\ArrowLine(37,-6)(67,-21)
\Line(67,-21)(82,-28)
\end{picture}}}
\newcommand{\cpbcp}{\bm{1in}{\begin{picture}(63,31) (35,-30)
\SetWidth{0.5}
\SetColor{Black}
\ArrowLine(97,-6)(67,-6)
\SetColor{SkyBlue}
\ArrowLine(67,-6)(37,-6)
\SetColor{Black}
\Vertex(67,-6){1.43}
\Vertex(52,-28){1.43}
\Vertex(82,-28){1.43}
\ArrowLine(82,-28)(97,-6)
\SetColor{SkyBlue}
\SetColor{Black}
\SetColor{Black}
\Vertex(37,-6){1.43}
\Vertex(97,-6){1.43}
\ArrowLine(37,-6)(67,-21)
\Line(67,-21)(82,-28)
\end{picture}}}
\newcommand{\cpaa}{\bm{1in}{\begin{picture}(63,31) (35,-30)
\SetWidth{0.5}
\SetColor{Black}
\ArrowLine(67,-6)(37,-6)
\Vertex(67,-6){1.43}
\Vertex(52,-28){1.43}
\Vertex(82,-28){1.43}
\ArrowLine(37,-6)(52,-28)
\ArrowLine(52,-28)(67,-6)
\SetColor{SkyBlue}
\SetColor{Black}
\SetColor{SkyBlue}
\SetColor{Black}
\Vertex(37,-6){1.43}
\Vertex(97,-6){1.43}
\end{picture}}}
\newcommand{\cpaap}{\bm{1in}{\begin{picture}(63,31) (35,-30)
\SetWidth{0.5}
\SetColor{SkyBlue}
\ArrowLine(67,-6)(37,-6)
\SetColor{Black}
\Vertex(67,-6){1.43}
\Vertex(52,-28){1.43}
\Vertex(82,-28){1.43}
\ArrowLine(37,-6)(52,-28)
\ArrowLine(52,-28)(67,-6)
\SetColor{SkyBlue}
\SetColor{Black}
\SetColor{SkyBlue}
\SetColor{Black}
\Vertex(37,-6){1.43}
\Vertex(97,-6){1.43}
\end{picture}}}

\newcommand{\cpab}{\bm{1in}{\begin{picture}(63,31) (35,-30)
\SetWidth{0.5}
\SetColor{Black}
\ArrowLine(97,-6)(67,-6)
\Vertex(67,-6){1.43}
\Vertex(52,-28){1.43}
\Vertex(82,-28){1.43}
\ArrowLine(82,-28)(97,-6)
\SetColor{SkyBlue}
\SetColor{Black}
\ArrowLine(67,-6)(82,-28)
\SetColor{SkyBlue}
\SetColor{Black}
\Vertex(37,-6){1.43}
\Vertex(97,-6){1.43}
\end{picture}}}
\newcommand{\cpabp}{\bm{1in}{\begin{picture}(63,31) (35,-30)
\SetWidth{0.5}
\SetColor{SkyBlue}
\ArrowLine(97,-6)(67,-6)
\SetColor{Black}
\Vertex(67,-6){1.43}
\Vertex(52,-28){1.43}
\Vertex(82,-28){1.43}
\ArrowLine(82,-28)(97,-6)
\SetColor{SkyBlue}
\SetColor{Black}
\ArrowLine(67,-6)(82,-28)
\SetColor{SkyBlue}
\SetColor{Black}
\Vertex(37,-6){1.43}
\Vertex(97,-6){1.43}
\end{picture}}}

\newcommand{\myfig}[3]{\begin{figure}[t]
\begin{center}
\epsfxsize=#2cm
\epsfbox{#1}
\end{center}
\caption{#3}
\label{fig:#1}
\end{figure}}

\title{Quantum Deformations from Toric Geometry}
\author{Samuel Pinansky\\ 
Department of Physics, UCSB, Santa Barbara, CA 93106\\
Email: \email{samuelp@physics.ucsb.edu}}
\abstract{We will demonstrate how calculations in toric geometry can be used to compute quantum corrections to the relations in the chiral ring for certain gauge theories.  We focus on the gauge theory of the del Pezzo 2, and derive the chiral ring relations and quantum deformations to the vacuum moduli space using Affleck-Dine-Seiberg superpotential arguments.  Then we calculate the versal deformation to the corresponding toric geometry using a method due to Altmann, and show that the result is equivalent to the deformation calculated using gauge theory.  In an appendix we will apply this technique to a few other examples.  This is a new method for understanding the infrared dynamics of certain quiver gauge theories.}
\preprint{hep-th/0511027}
\begin{document}
\section{Introduction}
An important problem that string theory purports to explain is the physics of singularities.  Singularities have been found to be both useful and natural to have in the compactification manifold of string theory, and have been studied for a long time.  One of the most important questions we wish to ask is what happens when fractional branes are stuck at singularities.  In the case of the conifold, we have a good understanding of what happens, however most singularities are more complicated and are not as well understood.
\par
Recently much progress has been made in expanding the class of singularities for which we can perform explicit calculations.  Using new geometric techniques infinite classes of theories have been constructed; first the $Y^{p,q}$ \cite{Martelli,Gauntlett,Benvenuti}, then the $X^{p,q}$ \cite{Kazakopoulos}, and finally the $L^{a,b,c}$ \cite{LpqrG,LpqrG2,LpqrF,LpqrM}.  These quiver gauge theories have moduli spaces which cover all cones over four sided polytopes, a particular subset of 3 complex dimensional toric varieties.  Both the gauge theory and the explicit metric on the gravity side of the AdS/CFT correspondence are known for the $Y^{p,q}$ and $L^{a,b,c}$ theories \cite{Martelli,Gauntlett,Benvenuti,LpqrG,LpqrF,franc,LpqrB}.  These examples have been used to perform careful tests of the AdS/CFT correspondence, and have also contributed to our understanding of the relationship between quiver gauge theories and other mathematical representations like toric geometry and dimer models.
\par 
In a similar line of work to deriving these new theories, new techniques have been developed for deriving conformal field theories and their properties directly from toric geometry using constructions like brane dimers \cite{GTFTG,BD}, Z-minimization \cite{ZminO,Zmin}, or more purely mathematical methods like exceptional collections \cite{ExCol,Bergman}.  Most of these methods have a direct geometric interpretation and (in the toric case) are easily turned into tractable computer algorithms.  Other methods using M-theory compactifications and deconstruction techniques are also being explored to analyze these theories and their quantum deformations \cite{Kap1,Kap2}.  However quantum corrections and their relationship to the classical toric geometry have not yet been fully explored using pure algebraic geometric techniques.
\par
For a long time, the only explicit solution we had of a quantum deformed singularity on both the gauge and gravity side was the conifold solution of Klebanov and Strassler \cite{KS,KW,MP,Kehag}.  However recent progress on Seiberg duality cascades and fractional branes \cite{HEK,chaotic,BHOP,FHSU,BBC} have shown heuristically that we can expect at least two different types of infrared behavior at the end of the cascades;  either supersymmetry is broken and the deformation is non-Calabi-Yau, or the deformation is similar to that of the conifold, or something else happens like the $SU(3)$ structure solutions of \cite{SU1}.  What exactly happens to the geometry when fractional branes corresponding to obstructed deformations are present is not well understood, and the infrared smoothed supergravity solutions are not known.  Even for the non-obstructed deformations, counting techniques using Minkowski decompositions or $p-q$ web splittings only gives a general picture of what we expect to happen and does not give us the mathematical tools to calculate the deformations exactly.  In this paper we will give a method due to Altmann \cite{Altmann}, using which can calculate the deformations to the geometry exactly from just the toric data, and we will show that this gives the correct result by comparing the calculation to the deformations of the relations in the chiral ring obtained from the gauge theory directly.  
\par
The previous work of \cite{BHOP} was inspired by a result in \cite{Altmann} that showed there were no unobstructed complex structure deformations for the cone over the del Pezzo 1.  Altmann's proof, however, is constructive, and we realized that it could be used to calculate deformations which were not obstructed purely from the toric geometric data.  Because the embedding relations of the toric geometry are classically exactly the relations among the chiral ring of the associated quiver gauge theory, it is natural to think that quantum deformations due to fractional branes which correspond to the non-obstructed deformations would be described by these deformations of the geometry as well.  In the case of the conifold, where the deformation and relations are trivial, the correspondence is clear, but for more complicated geometries that are not complete intersections the relationship remained to be worked out exactly.  In this paper I give the relationship between the deformations of the vacuum moduli space and the complex structure deformations calculated using Altmann's method for a particular non-trivial example, the cone over the del Pezzo 2, and find that they correspond exactly.  This example is complicated enough that we expect our results to hold more generally.
\par
The organization of the paper will be as follows: In section 2 we will introduce the gauge theory corresponding to the cone over the del Pezzo 2, and derive its chiral ring and the relations between the generators of the chiral ring using the F-term equations.  In section 3 we will find the ADS superpotential for the theory with $N$ fractional branes and $1$ probe brane, and use the corresponding F-term equations to solve for the relations of the deformed chiral ring.  In section 4, using the toric description of the geometry, we will derive two deformations, show that one is obstructed and one is free, and match the deformed geometry to the deformed relations in the chiral ring found in section 3.  Finally, in section 5 we will conclude by remarking how this method generalizes to other theories and commenting on some open problems for research.  Two appendices are also included. Appendix A contains a brief review of some of the toric geometry needed in this paper, as well as a calculation of the number of relations for the arbitrary $Y^{p,q}$.  Appendix B contains some further examples, including the deformed conifold as a simple illustration of our methods. 
\section{Classical Properties of the del Pezzo 2 theory}
The primary example for this paper will be the ${\cal N}=1$ quiver gauge theory corresponding to a cone over the second del Pezzo surface; i.e. the $dP_2$.  This theory is nice to work with because it has both deformation branes and supersymmetry breaking branes in the sense of \cite{BHOP,FHSU}.  It is also simple enough where the properties of the chiral ring can be analyzed directly on the gauge theory side.  This theory has already been explored from a number of perspectives.  For example, see \cite{FHH,ExColdP,MW,FHU} and references therein for various work that has been done.  In this section, we will define the theory, calculate the generators of the chiral ring, and find the relations among those generators that define the vacuum moduli space of the quiver theory.  
\subsection{Chiral primaries}
The gauge theory corresponding to the cone over the del Pezzo 2 is a quiver gauge theory with five gauge groups, with quiver diagram
\begin{equation}
\dpquiv,
\end{equation}
and superpotential:
\begin{align}
W&=\left(\cpbb\right)-\ \left(\cpaap\right)-\ \left(\cpabp\right)\label{superp}\\
\nonumber&+\ \left(\cpbap\right)+\ \left(\cpbcp\right)-\ \left(\cpdp\right).
\end{align}
We will use this graphical representation of the fields throughout this paper, where the traces are left implicit, and the symmetries of the equations are more readily apparent. 
We have colored three of the bifundamental fields in blue, because the F-term equations derived from the superpotential leave the number of blue fields invariant.  Therefore the number of blue fields in a loop provides a grading on the chiral primaries.
\par
To find the vacuum manifold of this theory, we place a single probe brane in the system, making the gauge group simply $U(1)^5$, so that all the bifundamentals commute.  Keeping the grading in mind, we can construct all loops with $0,1,2$, and $3$ blue fields, and then use the F-term equations to find a minimal independent set.  For loops with no blue fields we have, 
\begin{equation}
\left(\cpaa\right),\quad\left(\cpab\right),\quad\left(\quiv{\xl\zl\crr\zr}\right),\quad\left(\quiv{\zl\zr\cl\yr}\right).
\end{equation} 
But we also have the F-term equations for the field $\left(\quiv{\bb}\right)$ and $\left(\quiv{\ba}\right)$:
\begin{equation}
\left(\quiv{\zl\xl\crr}\right)=\left(\quiv{\xr\yr}\right),\quad\left(\quiv{\zr\yr\cl}\right)=\left(\quiv{\yl\xl}\right)
\end{equation}
so we only have two independent loops at this level.  We call them
\begin{equation}
\label{aeq}
a_1\equiv \left(\cpaa\right)\quad
a_2\equiv \left(\cpab\right).
\end{equation}
For the next level, we can repeat the argument.  There are $10$ different loops with one blue field, but the F-term equations show that only three of them are independent.  We call these three:
\begin{equation}
b_1\equiv \left(\cpba\right)\quad
b_2\equiv \left(\cpbb\right)\quad
b_3\equiv \left(\cpbc\right).
\end{equation}
With $2$ blue fields, there are $4$ different loops, but only $2$ of them are independent:
\begin{equation}
c_1\equiv \left(\cpca\right)\quad
c_2\equiv \left(\cpcb\right).
\end{equation}
And finally with all three blue fields there is only one loop, and we call it
\begin{equation}
\label{deq}
d\equiv \left(\cpd\right).
\end{equation}
\par
The 8 combinations of fields defined in (\ref{aeq})-(\ref{deq}) are the chiral primaries.  In other words, the generators of the chiral ring.  The manifold of their expectation values describes the vacuum manifold for the ${\cal N}=1$ supersymmetric quiver theory.  However this manifold is not simply $\mathbb{C}^8$, because these eight generators satisfy a number of relations between each other.
\subsection{Chiral ring relations}
These 8 chiral primaries are not independent; they satisfy a number of relations.  As an illustration we can show that $b_1b_3=b_2^2$:
\begin{align}
\nonumber\left(\cpba\right)\cdot\left(\cpbc\right)&=\left(\quiv{\ba\xl\yl}\right)\cdot\left(\quiv{\bb\xr\yr}\right)\\
&=\left(\cpbb\right)\cdot\left(\cpbb\right)
\end{align}
where in the first equation we used the F-term equation for the field $\left(\quiv{\bc}\right)$, and the second equation used the F-term equations for the field $\left(\quiv{\xl}\right)$ and $\left(\quiv{\yr}\right)$.
\par
After doing many calculations like this, we find that there are 14 relations:
\begin{align}
\nonumber b_2^2=b_1b_3,\quad
b_2^2=a_1c_2,\quad
b_2^2=c_1a_2,\quad
c_1^2=b_1d,\quad
c_2^2=b_3d\\
\nonumber b_1a_2=b_2a_1,\quad
c_1b_2=c_2b_1,\quad
b_2a_2=b_3a_1,\quad
c_1b_3=c_2b_2\\
b_1b_2=c_1a_1,\quad
b_2b_3=c_2a_2,\quad
c_1c_2=b_2d,\quad
c_1b_2=a_1d,\quad
c_2b_2=a_2d.
\label{delP2rel}
\end{align}
These $8$ complex variables and $14$ relations are not a complete intersection in $\mathbb{C}^8$, because the dimension of the variety, $3$, is greater than $m-n$, where $m=8$ is the number of variables and $n=14$ is the number of equations. So therefore, in some sense, this representation of the space is very tightly constrained.  It defines a cone over the second del Pezzo surface, or $dP_2$.  
Because this corresponds to a toric geometry, the chiral primaries can be arranged to make the relationships between them correspond to linear relations between vectors:
\begin{align}
\nonumber\begin{array}{cccc}
\quad\quad&\quad\quad&b_3& \\
 &c_2&\quad\quad&a_2\\
d&\quad\quad&b_2& \\
 &c_1&\quad\quad&a_1\\
 &\quad\quad&b_1& 
\end{array}
\end{align}
As is explained in Appendix A, the relations correspond to irreducible parallelograms in the above diagram, and it is straightforward to verify that those relations are exactly the ones found from the F-term equations above, (\ref{delP2rel}).  
\section{Quantum Deformations of $dP_2$}
When all five gauge groups have the same rank, this theory is conformal.  However we want to study the behavior of the non-conformal theory with fractional branes, specifically the IR behavior where we expect some of the gauge groups to confine.  In this section we will find the deformed vacuum moduli space in the IR by solving for the ADS deformed superpotential when you place $N$ deformation fractional branes on the quiver.  Then we will use that solution to analyze what happens to the chiral ring relations we derived above.
\subsection{Branes on $dP_2$}
We can analyze the various branes we can put on this quiver theory. Ordering the gauge groups as
\begin{equation}
\nonumber \quiv{\Text(67,0)[lb]{\normalsize{\Black{$3$}}}
\Text(52,-22)[lb]{\normalsize{\Black{$2$}}}
\Text(82,-22)[lb]{\normalsize{\Black{$4$}}}
\Text(37,0)[lb]{\normalsize{\Black{$1$}}}
\Text(97,0)[lb]{\normalsize{\Black{$5$}}}}
\end{equation}
we have a bulk brane with weight $(1,1,1,1,1)$, a fractional brane with weight $(1,2,1,0,0)$, and another fractional brane with weight $(1,0,0,0,1)$.  We expect to get two different fractional branes because the toric polytope for $dP_2$ has $5$ sides (equivalently, the $p-q$ web has five legs), as will be explored in section 4.  Both fractional branes are expected to trigger a duality cascade \cite{HEK}.  However the first brane will end with supersymmetry breaking (the so-called Supersymmetry Breaking by Obstructed Geometry, or SUSY BOG \cite{BHOP}), while the second brane will trigger a complex structure deformation (the so-called deformation brane \cite{FHSU}).  First we review what happens in the case where we have $N$ supersymmetry breaking branes, and show that the deformation is incompatible with the F-term equations.
\par
Placing $N$ SUSY BOG branes leaves us with this quiver diagram:
\begin{equation}
\bm{.9in}{\begin{picture}(63,51)(35,-40)\SetWidth{0.5}\SetColor{Black}
\Vertex(67,-6){1.43}
\Vertex(52,-28){1.43}
\Vertex(37,-6){1.43}
\xl\yl\zl
\SetColor{SkyBlue}
\ArrowArc(52,-17)(18,36.87,143.13)
\Text(67,0)[lb]{\normalsize{\Black{$SU(N)$}}}
\Text(52,-42)[lb]{\normalsize{\Black{$SU(2N)$}}}
\Text(0,0)[lb]{\normalsize{\Black{$SU(N)$}}}
\end{picture}}
\end{equation}
and simply the second term in the superpotential given in (\ref{superp}).
We expect that these gauge groups will flow to strong coupling in the infrared, but using the same arguments that were used in \cite{BHOP}, we can use the Konishi anomaly equations \cite{Konishi, CDSW}:
\begin{equation}
\left\langle -\frac{1}{32\pi^2}
\sum_{i,j} \left[ W_\alpha, \left[W^\alpha, 
\frac{\partial f} {\partial {X_i^j}} \right] \right] \right \rangle = 
\left \langle \tr  (f(X,W_\alpha W^\alpha) \partial_X W) \right \rangle \ .
\end{equation}
where $\delta X=f(X,W_\alpha W^\alpha)$ is the variation of the bifundamental $X$, and $W_\alpha$ is the supersymmetric gauge field strength.  If we take the simplest variation, $\delta X=X$ for the $4$ remaining fields, we get
\begin{align}
\tr\left\langle
\bm{.5in}{\begin{picture}(63,51)(35,-40)\SetWidth{0.5}\SetColor{Black}
\Vertex(67,-6){1.43}
\Vertex(52,-28){1.43}
\Vertex(37,-6){1.43}
\xl\yl\ba
\end{picture}}
\right\rangle&=N S_1+ NS_3=2N S_1+NS_2= 2N S_3+ N S_2\\
0&=N S_1+N S_3
\end{align}
as the equations for each field, where we have defined $S_k=-\tr_k W_\alpha W^\alpha/32\pi^2$.  The only solution to these equations is $S_1=S_2=S_3=0$, which implies that something other than a deformation of the solutions to the F-term equations happens, as discussed in detail in \cite{BHOP, FHSU, BBC}.
\par
We will focus for the rest of the paper on what happens when you have $N$ deformation branes, $1$ probe brane, and no SUSY BOG causing branes.
 \subsection{ADS superpotential}
We wish to study the conifold-like deformations of the complex structure of the $dP_2$ gauge theory, so we now turn to the case where only probe branes and deformation branes are present, where we expect the deformation is {\em not} obstructed, and we can fully analyze the geometry by looking at the chiral ring relations.
\par
If we add $N$ deformation branes and $1$ probe brane, the ranks of the gauge groups will be $(N+1,1,1,1,N+1)$.  Since the $SU(N+1)$ groups only have bifundamentals connected to $U(1)$ groups, we can argue that the behavior in the infrared for nodes $1$ and $5$ is exactly like a normal $SU(N+1)$ theory with 2 flavors, one for each bifundamental connected to the node.  Under the assumption that these gauge groups confine, we are told to rewrite the superpotential in terms of combinations of fields which are gauge invariant with respect to the confining gauge groups.
\par
We call these four combinations:
\begin{align}
\nonumber m^{11}\equiv \left(\quiv{\ba\xl}\right),\quad&\quad m^{12} \equiv \left(\quiv{\ba\cl}\right)\\
m^{21}\equiv \left(\quiv{\zl\xl}\right),\quad&\quad m^{22}\equiv \left(\quiv{\zl\cl}\right)
\end{align}
and likewise for the confining gauge group on the right with the variables $n^{ab}$, and also give the 3 extra singlet fields these labels:
\begin{align}
l\equiv \left(\quiv{\yl}\right),\quad\quad
r\equiv \left(\quiv{\xr}\right),\quad\quad
b\equiv \left(\quiv{\bc}\right).
\end{align}
Now we can rewrite the superpotential in terms of these commuting fields:
\begin{equation}
W=lrb-m^{11} l-n^{11} r+m^{21} n^{12}+m^{12} n^{21}-m^{22}n^{22} b.
\end{equation}
In the infrared, the superpotential develops an ADS piece like
\begin{equation}
W_{ADS}=S_1\log(\det(m))+S_5\log(\det(n))
\end{equation}
where $S_1$ and $S_5$ are the gaugino condensates for the first and fifth gauge groups.  Now we can find the correct infrared moduli space by looking at the solutions to the F-term equations, $\partial (W+W_{ADS})/\partial X=0$, for $X$ each of the $11$ variables, $m^{ab}, n^{ab}, l, r,$ and $b$.  
The F-term equations for $l,r$, and $b$ give that
\begin{equation}
m^{11}=rb,\quad n^{11}=lb,\quad m^{22}n^{22}=lr
\label{ftermbas}
\end{equation}
The equations for the off diagonal $m$ and $n$ variables give
\begin{equation}
S_1S_5=\det(m)\det(n),\quad m^{12}n^{21}=m^{21}n^{12}.
\end{equation}
And the equations for the diagonal $m$ and $n$ variables gives that
\begin{equation}
lm^{21}=m^{22}n^{21},\quad rn^{21}=n^{22}m^{21},\quad lm^{12}=m^{22}n^{12},\quad rn^{12}=n^{22}m^{12}.
\label{ftermsolutions}
\end{equation}
Finally, we can solve for $S_1$ and $S_5$ given the above equations, giving
\begin{equation}
S\equiv S_1=S_5=lrb-m^{12}n^{21}.
\label{ftermsolutionsderived}
\end{equation}
The chiral primaries in terms of our new fields are:
\begin{align}
\nonumber a_1=m^{21}l,\quad a_2&=n^{21}r\\
\nonumber b_1=m^{22}n^{11},\quad b_2=blr,\quad b_3&=m^{11}n^{22}\\
\nonumber c_1=m^{12}n^{11},\quad c_2&=m^{11}n^{12}\\
d&=m^{12}n^{12}b.
\end{align}
\subsection{Deformed chiral ring}
Now that we know how the F-term equations are modified by the quantum effects, we want to see what happens to the 14 relations among the chiral ring (\ref{delP2rel}).  First we look at the relation $db_1=c_1^2$.  We use the solutions to the F-term equations to prove or disprove this relation:
\begin{align}
\nonumber db_1&=m^{12}n^{12}bm^{22}n^{11}\\
\nonumber &=m^{12}m^{12}n^{11}n^{11}\\
&=c_1^2,
\end{align}
where we used (\ref{ftermbas}) and (\ref{ftermsolutions}). 
\par
As an example of a relation which is modified, we examine $c_1 a_2=b_2^2$:
\begin{align}
\nonumber c_1a_2&=m^{12}n^{11}n^{21}r\\
\nonumber &=(blr-S)blr\\
&=b_2(b_2-S),
\end{align}
where we used (\ref{ftermbas}) and (\ref{ftermsolutionsderived}).
We proceed in this manner through the other 12 equations.  Summarizing the results, we find that the 14 relations among the chiral primaries are deformed to:
\begin{align}
\nonumber b_2^2=b_1b_3,\quad b_2(b_2-S)=a_1 c_2,\quad b_2(b_2-S)= c_1 a_2,\quad b_1 a_2&=b_2 a_1,\\
\nonumber c_1 b_2=c_2 b_1,\quad b_2 a_2 = b_3 a_1,\quad c_1 b_3=c_2 b_2,\quad b_1(b_2-S)&= c_1 a_1,\\
\nonumber (b_2-S) b_3= c_2 a_2,\quad c_1 c_2=b_2 d,\quad c_1(b_2-S)=a_1 d,\quad c_2 (b_2-S)&=a_2 d,\\
 c_1^2=b_1 d,\quad c_2^2&=b_3 d\label{deformedfterm}.
\end{align}
In the next section, we will derive this exact deformation by using the toric geometry techniques of Altmann.
\section{Deformations of $dP_2$ from Toric Geometry}
The moduli space of this quiver theory has an alternate description in terms of toric geometry.  In this section we will use this description to derive the deformed space (\ref{deformedfterm}) using only the combinatoric data defining the toric geometry.  We will first review the construction of the original, undeformed moduli space, and then proceed to construct the necessary data to find the versal deformation of the toric singularity, using the methods of Altmann \cite{Altmann}.
\subsection{14 relations that define $dP_2$}
\myfig{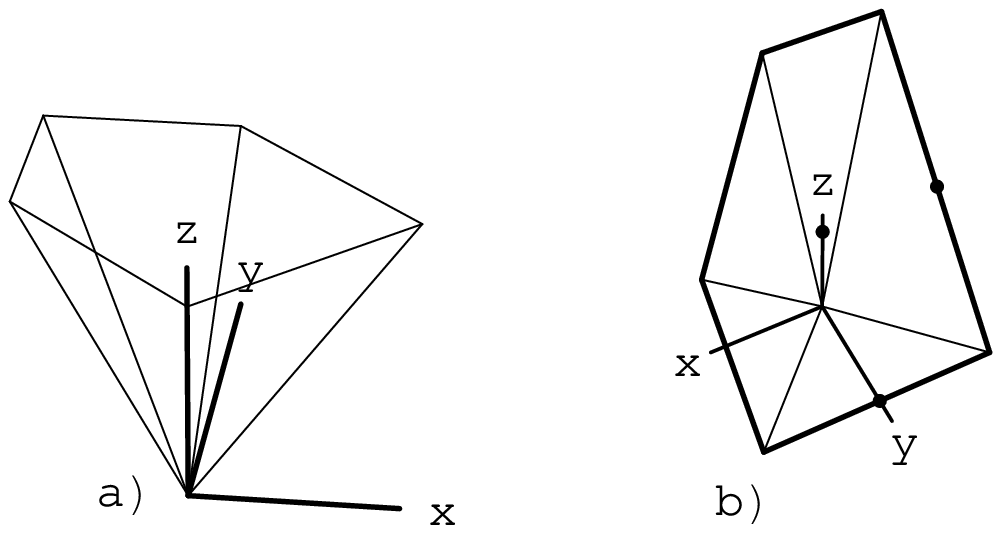}{12}{(a) The lattice cone defining the cone over the del Pezzo 2, (b) The dual lattice cone to the cone over the del Pezzo 2.  The marked points are the extra generators needed to generate the lattice cone.}
The toric variety of the cone over the del Pezzo 2 can be defined by five vectors:
\begin{equation}
v_1=(0,0,1),\quad v_2=(1,1,1),\quad v_3=(0,2,1),\quad v_4=(-1,2,1),\quad v_5=(-1,1,1)
\end{equation}
which define the cone in the lattice shown in figure 1(a).
We can describe the geometry as an intersection in a higher complex dimensional space by constructing the minimal set of generators of the dual cone, i.e. the cone over the lattice defined by the inward pointing normals to this cone (see Appendix A for a brief review of the toric geometry techniques).   
Figure 1b shows the dual cone; it is defined by the vectors:
\begin{equation}
\tilde v_1=(-1,1,0),\quad \tilde v_2=(-1,-1,2),\quad \tilde v_3=(0,-1,2),\quad \tilde v_4=(1,0,1),\quad \tilde v_5=(1,1,0).
\end{equation}
However, these five vectors by themselves do not generate the entire lattice cone.  We must add the three extra lattice vectors that lie within the base of the cone:
\begin{equation}
\tilde v_6=(0,1,0),\quad \tilde v_7=(-1,0,1),\quad \tilde v_8=(0,0,1).
\end{equation}
The geometry can then be defined as an intersection in ${\mathbb{C}^8}$, by associating to each of these generating vectors a complex coordinate subject to constraints defined by the linear relationships between the generating vectors.  We call these 8 coordinates (suggestively) $a_1,a_2,b_1,b_2,b_3,c_1,c_2,$ and $d$, where:
\begin{align}
\nonumber  a_1\sim\tilde v_3,\quad a_2\sim \tilde v_4,\quad b_1&\sim \tilde v_2,\\
\nonumber  b_2\sim \tilde v_8,\quad b _3 \sim\tilde v_5,\quad c_1&\sim\tilde v_7,\\
c_2\sim \tilde v_6,\quad d&\sim\tilde v_1.
 \end{align}
 Now a linear relationship between the $\tilde v$'s defines a polynomial relationship between the $a,b,c,d$ embedding coordinates.  For example, $\tilde v_8+\tilde v_8=\tilde v_2+\tilde v_5$ means that $b_2^2=b_1 b_3$.  A minimal set of these relations can be found by counting all irreducible (in the sense described in Appendix A) parallelograms in the face of the dual cone, including degenerate ones that reduce to a line, as in the example $b_2^2=b_1 b_3$.  It is straightforward to count that there are 14 such parallelograms.  They give rise exactly to the 14 relations in the chiral ring for the $dP_2$ quiver theory we analyzed in section 2, as of course it must (otherwise we choose the wrong quiver and/or superpotential for the original theory):
 \begin{align}
b_2^2=b_1b_3,\quad
b_2^2=a_1c_2,\quad
b_2^2=c_1a_2,\quad
c_1^2=b_1d,\quad
c_2^2&=b_3d\\
\nonumber b_1a_2=b_2a_1,\quad
c_1b_2=c_2b_1,\quad
b_2a_2=b_3a_1,\quad
c_1b_3&=c_2b_2\\
\nonumber b_1b_2=c_1a_1,\quad
b_2b_3=c_2a_2,\quad
c_1c_2=b_2d,\quad
c_1b_2=a_1d,\quad
c_2b_2&=a_2d
\end{align}
\subsection{Altmann's Construction of the Space of Versal Deformations}
A mathematical problem we might want to solve can be stated this way:  Given an intersection (not necessarily complete) of hypersurfaces inside of $\mathbb{C}^{n}$, how can we modify the equations and maintain both the dimension and the degree of the algebraic variety\footnote{The appropriate mathematical definition here is flatness, or the consistency of all the Hilbert polynomials along the family of varieties.  The dimensionality of the manifold is only one possible invariant.}?  Two examples are explored very thoroughly in this paper.  The first example is the conifold, given by the equation $uv=wz$ imbedded in $\mathbb{C}^4$.  In this case the answer to our question is simple; there is only one thing you can do, namely add a constant $uv=wz+\epsilon$.  The second example is the cone over the del Pezzo 2, given by the 14 relations in (\ref{delP2rel}).  Unlike the conifold, simply adding an $\epsilon$ to one of the relations fails, reducing the solution set to dimension 0, which clearly will not correspond to a physical quantum deformation of the space.
\par
Altmann \cite{Altmann} gives a method of finding all such deformations for the specific case when the intersection is a toric Gorenstein singularity.  He proves that
\begin{theorem}
The space of deformations of the complex structure of a toric Gorenstein singularity can be found by first lifting the defining relations to the tautological cone, and then restricting to the ideal of the base of the cone over the Minkowski summands.
\end{theorem}
\par
In the rest of this section we will define those various cones, and explain how one can apply this in practice.  We will not go into the details of the proof of this statement; the full proof can be found in \cite{Altmann}.  The content of the theorem can be summarized like this:  By lifting the defining relations to a higher dimensional space in a certain way, we capture the most generic possible deformation of the relations.  However the extra coordinates in this higher dimensional space are not independent.  They also satisfy relations determined by the original geometry.  Therefore we must also find these relations and restrict to them.  The process consists of two steps:  First, lift the relations to the higher dimensional space.  Second, restrict these new dimensions due to the relations between the new coordinates.
\par
First we define the cone over the Minkowski summands, which I will simply refer to as the Minkowski cone.  This cone defines the extra dimensions we will add to the original geometry.  Let $Q$ be the original 2 dimensional lattice polytope and let $\sigma(Q)$ be the lattice cone with that polytope as its base, as set up in Appendix A.  Choose one vertex of $Q$ and place it at $(0,0)$.  Now we can describe $Q$ as a sequence of edges, $d^i$, for $i=1..N$, where $N$ is the number of sides of $Q$.  For example, for the conifold:
\begin{equation}
d=\{(1,0),(0,1),(-1,0),(0,-1)\}
\end{equation}
where we have placed the lower left corner at the origin.  This gives us a translationally invariant way of describing the polytope.  For the del Pezzo 2, as shown here:
\begin{equation}
\bm{1in}{\begin{picture}(66,80) (56,-74)
\SetWidth{0.5}
\SetColor{Black}
\Vertex(60,6){2.83}
\Vertex(90,6){2.83}
\Vertex(60,-24){2.83}
\Vertex(90,-54){2.83}
\Vertex(120,-24){2.83}
\Line(90,6)(60,6)
\Line(60,6)(60,-24)
\Line(60,-24)(90,-54)
\Line(90,-54)(120,-24)
\Line(120,-24)(90,6)
\Vertex(60,-54){2.83}
\Vertex(120,-54){2.83}
\Vertex(120,6){2.83}
\Vertex(90,-24){2.83}
\Text(80,-74)[lb]{\normalsize{\Black{$(0,0)$}}}
\end{picture}}
\label{dP2poly}
\end{equation}
we can define it by a series of vectors along its face:
\begin{equation}
d=\{(1,1),(-1,1),(-1,0),(0,-1),(1,-1)\}.
\end{equation}
The Minkowski cone, $C(Q)$,  is then defined as
\begin{equation}
C(Q)\equiv V\cap {\mathbb{R}^N_{\geq 0}}\quad{\rm with}\quad V\equiv\{t_1,\dots,t_N\;|\; t_id^i=0\}.
\end{equation}
\myfig{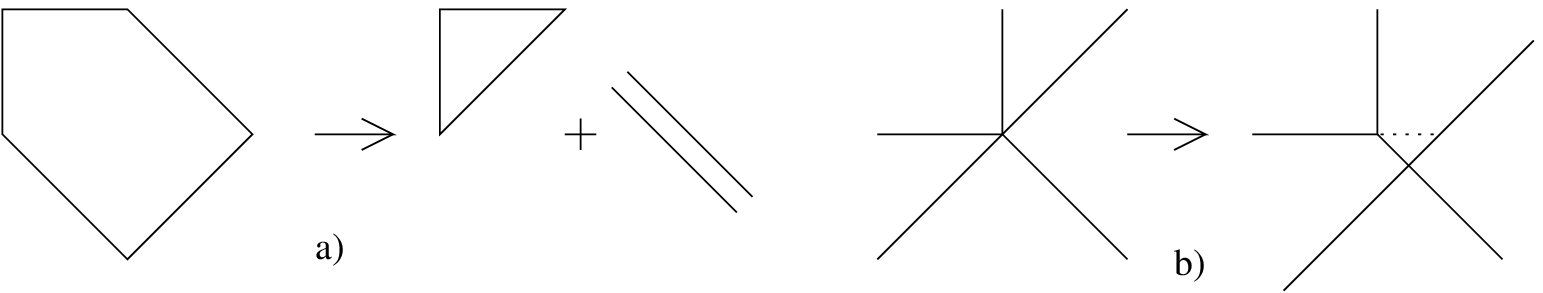}{12}{(a) Splitting the polytope into two Minkowski summands (b) Separating the p-q web}
For the conifold, we can see that $C(Q)={\mathbb{R}^2_{\geq 0}}$.  Every point in this rational cone defines a Minkowski summand of the original polytope.  Specifically, each point $t\in C(Q)$ defines a polytope $Q_t$ defined by the edges $t_id^i$ (no sum).  For the case of the $dP_2$, we can visualize this in terms of separating the original polytope into two sub-polytopes, or in terms of p-q web splitting (see figure 2).
\par
The base of the Minkowski cone defines an ideal, given by the vector polynomials
\begin{equation}
\sum_{i=1}^N t_i^kd^i
\end{equation}
for each $k\geq1$.  However, for $k>w$, where $w$ is the smallest integer such that the polytope $Q$ can be contained in 2 linearly independent strips of width $w$, we can drop those equations as they are generated by the ones of lesser degree.  This ideal is the restriction on the extra dimensions we are going to add to the original geometry.  For our conifold example the width is simply $1$ since it's a unit square.  
Therefore the ideal is generated by two linear polynomials:
\begin{equation}
t_1-t_3,\quad t_2-t_4.
\end{equation}
For the $dP_2$ the width is $2$ since there is only one strip of width one which the polytope fits in. Therefore we need to keep the quadratic equations, and the ideal is generated by
\begin{align}
\nonumber t_1-t_2-t_3+t_5&=0\\
\nonumber t_1^2-t^2_2-t^2_3+t_5^2&=0\\
\nonumber t_1+t_2-t_4-t_5&=0\\
t_1^2+t_2^2-t_4^2-t_5^2&=0.
\label{minkbase}
\end{align}
We can parameterize the solutions to the linear constraints with three variables $t, s_1, s_2$, such that $t_1=t,\; t_2=t-s_1,\; t_3=t-s_2,\; t_4=t+s_2$, and $t_5=t-s_1-s_2$.  Then the quadratic constraints become
\begin{align}
\nonumber 2s_1s_2&=0\\
s_2^2&=0
\end{align}
forcing $s_2=0$.  So there is only one free parameter, $s_1$, which is the deformation caused by the deformation fractional brane.  We interpret this parameter as the gluino condensate field $S$ on the field theory side.  We will see that this interpretation exactly matches with the results calculated in section 3.
\par
Now that we have a way of describing how to split up the polytope into Minkowski summands, we would like to combine that information with the original cone.  This combined cone is called the ``tautological'' cone, and we denote it as $\widetilde{C}(Q)$.  The tautological cone is the space combining the original geometry with the extra dimensions corresponding to possible deformations.  Recall that the original cone over the polytope is $\sigma(Q)\subseteq\mathbb{R}^3$ and that $C(Q)\subseteq V\cap \mathbb{R}^N_{\geq 0}$.  We want to form a cone which combines these two so that $\widetilde{C}(Q)\subseteq \mathbb{R}^3\times V$.  The challenge is to find some consistent way of lifting from the original cone $\sigma(Q)$ to the tautological cone $\widetilde{C}(Q)$.
\par
Since we have already placed one vertex of $Q$ at $(0,0)$, we can define every vertex $a\in Q$ in terms of a path $\lambda^i$ such that $a=\lambda_id^i$.  Clearly, $\lambda^i\in\mathbb{Z}^N$ for lattice polytopes.  The choice of $\lambda$ is not unique, however it can be shown that different choices of $\lambda$ end up giving the same result \cite{Altmann}.  Just as we defined a point in the polytope $Q$, we can define a point in the Minkowski summand $Q_t$ as $a_t=t_i\lambda_id^i$.  Now we see how to define the tautological cone:
\begin{equation}
\widetilde{C}(Q)\equiv\{(a_t,t)\;|\; t\in C(Q), a_t\in Q_t\}.
\end{equation}
To take this lifting and dualize it (i.e. apply it to the defining relations directly), we take the generator corresponding to $(0,0,1)$\footnote{This generator will always be present for any polytope with at least one interior lattice point.  See Appendix B.1 for how to modify the algorithm in the special case when it is not.} and replace it by the $t_i$'s according to some power weights. For each generator $\tilde v_i$ except for $(0,0,1)$, we split it up into a 2-vector and a scalar: $\tilde v_i=(c_i,\eta_0(c_i))$.  Then we find a point $a(c_i)\in Q$ such that $-c_i\cdot a(c_i)=\eta_0(c_i)$.  Then we find the path representation for the $a(c_i)$, i.e. find $\lambda_j^{c_i}$ such that $a(c_i)=\lambda_jd^j$.  Finally, we can define the weights for lifting to the tautological cone: 
\begin{equation}
\eta(c_i)\equiv (-\lambda_1^{c_i}(d^1\cdot c_i),\dots,-\lambda_N^{c_i}(d^N\cdot c_i))\in \mathbb{Z}^N.
\end{equation}
Now if we call $z_i$ the complex embedding coordinate correspoding to the generator $\tilde v_i$, and call the coordinate corresponding to the $(0,0,1)$ generator $t$, we replace the $t$ by the $N$ Minkowski cone variables $t_1,t_2,\dots,t_N$ in the defining embedding relations according to the power weights given by the $\eta(c_i)$ defined above.  In general, a relation of the form 
\begin{equation}
\prod_i t^az_i^{p_i}=\prod_i z_i^{q_i}
\end{equation}
becomes:
\begin{equation}
\prod_i t_i^{\left(\sum_j p_j\eta(c_j)-q_j\eta(c_j)\right)_i}z_i^{p_i}=\prod_i z_i^{q_i}.
\end{equation}
Note that due to this construction, it can be shown that $a=\sum_i\left(\sum_j p_j\eta(c_j)-q_j\eta(c_j)\right)_i$.
\par
We now follow this procedure for the $dP_2$.  We have already found the generators of the dual cone in section 4.1:
\begin{align}
\nonumber \tilde v_1=(-1,1,0),\quad \tilde v_2=(-1,-1,2),\quad \tilde v_3=(0,-1,2),\quad \tilde v_4=(1,0,1),\quad \tilde v_5=(1,1,0)\\
\tilde v_6=(0,1,0),\quad \tilde v_7=(-1,0,1),\quad \tilde v_8=(0,0,1).
\end{align}
Splitting them into $(c^i,\eta_0(c^i))$, and then finding points on the polytope $a(c^i)$ such that $-c_i\cdot a(c_i)=\eta_0(c_i)$, we get:
\begin{align}
\nonumber c^1=(-1,1),\quad a(c^1)=(0,0),\quad \eta_0(c^1)&=0\\
\nonumber c^2=(-1,-1),\quad a(c^3)=(1,1),\quad \eta_0(c^2)&=2\\
\nonumber c^3=(0,-1),\quad a(c^4)=(-1,2),\quad \eta_0(c^3)&=2\\
\nonumber c^4=(1,0),\quad a(c^5)=(-1,2),\quad \eta_0(c^4)&=1\\
\nonumber c^5=(1,1),\quad a(c^6)=(0,0),\quad \eta_0(c^5)&=0\\
\nonumber c^6=(0,1),\quad a(c^7)=(0,0),\quad \eta_0(c^6)&=0\\
c^7=(-1,0),\quad a(c^2)=(1,1),\quad \eta_0(c^7)&=1.
\end{align}
If we define $\lambda^i$ as vector paths to the points $a(c^i)$, i.e. $\sum \lambda^i_j d^j=a(c^i)$, we can choose
\begin{align}
\nonumber \lambda^1&=\lambda^5=\lambda^6=(0,0,0,0,0)\\
\nonumber \lambda^2&=\lambda^7=(1,0,0,0,0)\\
\lambda^3&=\lambda^4=(0,1,0,-1,0).
\end{align}
Then we can define the weight vectors $\eta(c^i)$, as $(-\lambda_1^{c_i}(d^1\cdot c_i),\dots,-\lambda_5^{c_i}(d^5\cdot c_i))$, giving:
\begin{align}
\nonumber \eta(c^1)&=\eta(c^5)=\eta(c^6)=(0,0,0,0,0)\\
\nonumber \eta(c^2)&=(2,0,0,0,0)\\
\nonumber \eta(c^3)&=(1,1,0,0,0)\\
\nonumber \eta(c^4)&=(0,1,0,0,0)\\
\eta(c^7)&=(1,0,0,0,0).
\end{align}
These define how to lift the variables to a $7+5=12$ dimensional dual tautological cone.  If we relabel to match our previous notation, $c^1\to d,\; c^2\to b_1,\; c^3\to a_1,\; c^4 \to a_2,\; c^5\to b_3,\; c^6\to c_2,\; c^7\to c_1$, we can take our $14$ defining equations and lift them to the higher space by replacing $b_2$, the generator which corresponds to $(0,0,1)$, with five variables, $t_1,\dots, t_5$ according to the power weight vectors $\eta$.  For example, the equation 
\begin{equation}
b_2^2=b_1b_3
\end{equation}
becomes:
\begin{equation}
t_1^2=b_1 b_3,
\end{equation}
since $\eta(b_1)+\eta(b_3)=(2,0,0,0,0)$.  Proceeding through the rest of the equations, we find that they become:
\begin{align}
\nonumber t_1^2=b_1b_3,\quad t_1t_2=a_1 c_2,\quad t_1t_2= c_1 a_2,\quad b_1 a_2&=t_1 a_1,\\
\nonumber c_1 t_1=c_2 b_1,\quad t_1 a_2 = b_3 a_1,\quad c_1 b_3=c_2 t_1,\quad b_1 t_2&= c_1 a_1,\\
\nonumber t_2 b_3= c_2 a_2,\quad c_1 c_2=t_1 d,\quad c_1 t_2=a_1 d,\quad c_2 t_2&=a_2 d,\\
c_1^2=b_1 d,\quad c_2^2&=b_3 d.
\end{align}
\par
According to the theorem, the final step is to restrict these equations to the ideal over the base of the Minkowski cone.  So we simply parameterize the solutions to the ideal, and plug those parameters into the lifted equations.  If we only looked at the 2 linear relations, we see that the $t$ parameter is replaced by $N-2$ coordinates, with a net gain of $N-3$ degrees of freedom.  Therefore we expect there to be $N-3$ different fractional branes for a quiver theory that corresponds to an $N$ sided polytope.  However further restrictions on deformation can arise at the quadratic and higher order ideal polynomials.  In general, it seems to be easiest to first parameterize the solutions to the linear equations, and then use the higher order equations in order to derive further restrictions on the parameters.
\par
For the $dP_2$ the five variables are restricted to lie on the base of the cone over the Minkowski summands on the ideal generated by the solutions to (\ref{minkbase}).
Substituting these definitions back into the equations, we arrive at the equations defining the deformed moduli space (and relabeling $t\to b_2$ and $s_1\to S$)
\begin{align}
\nonumber b_2^2=b_1b_3,\quad b_2(b_2-S)=a_1 c_2,\quad b_2(b_2-S)= c_1 a_2,\quad b_1 a_2&=b_2 a_1,\\
\nonumber c_1 b_2=c_2 b_1,\quad b_2 a_2 = b_3 a_1,\quad c_1 b_3=c_2 b_2,\quad b_1(b_2-S)&= c_1 a_1,\\
\nonumber (b_2-S) b_3= c_2 a_2,\quad c_1 c_2=b_2 d,\quad c_1(b_2-S)=a_1 d,\quad c_2 (b_2-S)&=a_2 d,\\
c_1^2=b_1 d,\quad c_2^2&=b_3 d.
\end{align}
Comparing these equations with (\ref{deformedfterm}), we find exact agreement.  
\par
This exact agreement is actually quite surprising.  Although we expected the geometric method to find the same deformed geometry, we did not expect to find it in the same coordinates.  Indeed, had we chosen a different parameterization for the solution to the ideal (\ref{minkbase}), we would have found a different set of deformed equations, equivalent through some change of variables in the embedding coordinates.  The surprising fact that the ``simplest'' choice of parameterization for both the gauge theory and the geometric method coincide is most likely due to the fact that in this example there is only one non-obstructed deformation.  In theories such as the del Pezzo 3, with more than one deformation, the deformations may mix with each other, and we expect that one would have to work harder to show the equivalence between the deformed geometry and the deformed chiral ring.
\section{Conclusions}
In this paper we have calculated the chiral primaries and the relations between them in two settings, through the gauge theory using ADS superpotential techniques, and through the geometry side using toric techniques.  The latter has the advantage of being more easily programmable on a computer, and only uses the combinatoric structure of the toric singularity.  While up until recently, these geometric methods were only useful for deriving classical moduli spaces, we have shown in a specific calculation that they can also be used to analyze the quantum behavior of moduli spaces, opening the possibility of gaining a better understanding of the quantum deformations of toric singularities of all types.
\par
Much work remains to be done, however.  As the del Pezzo 2 corresponds to the $X^{2,1}$ space \cite{Kazakopoulos}, whether this analysis generalizes to the more complicated geometries of the $X^{p,q}$ spaces is an obvious thing to explore.  An interesting question is whether you can use these solutions to help construct the corresponding deformed supergravity solution in the spirit of the deformed conifold, as the connection between these geometric techniques and supergravity is still not well understood and exploring it may shed some light on the more general AdS/CFT correspondence.  An important open problem is to understand what happens if we add deformation obstructed branes, breaking supersymmetry through the SUSY BOG mechanism, and if this formalism might be generalized to give us some way of describing what happens to the geometry in those situations.  If we have a better quantitative understanding of how these geometries deform, we may be able to approach the issues involved in compactifying them, which opens up a number of possible phenomenological uses for these theories. This is just one more of a set of powerful new tools with which we can calculate explicit geometries of non-trivial gauge theories.
\section*{ Acknowledgments }
I am very grateful to my advisor D. Berenstein and to A. Hanany, C. Herzog, P. Ouyang, and D. Vegh for many discussions related to this work. Work supported by the DOE, under grant DE-FG02-91ER40618.
\appendix
\section{Some Techniques from Toric Geometry}
In this appendix we will review how to analyze the toric description of these types of geometries, and how to go from a description of the singularity as a cone over a lattice polytope to the embedding as an intersection in some higher complex dimensional space.
\par
Theorem 4.1 applies to a class of varieties that are cones over toric Gorenstein singularities.  This means that they can be represented by lattice cones with polytopes as their base, where no lattice points lie along the edges of the base.  The case of most interest to physicists is the case of a polygon in a 2 dimensional lattice, because the cone over that geometry is a 3 complex dimensional space.  3 dimensional polytopes might also be useful for F-theory compactifications; however for simplicity I will specialize to 2 dimensions for this paper.   The interested reader can refer to \cite{Altmann} for details on how to modify these calculations for arbitrary dimension.
\par
Given a lattice polygon, a basic question we would like to answer is ``How do we take this description and recover the description of the geometry in terms of an intersection (not necessarily complete) of hypersurfaces?''  To do this is a straightforward three step process:
\begin{enumerate}
\item
Form the lattice cone with the polytope as its base.
\item
Find the minimal set of generators of the dual lattice cone.
\item
Associate to each generator a complex variable. Then the linear relations among the generators define the multiplicative relations among the complex variables which define your intersection.
\end{enumerate}
As a trivial example we can examine the conifold, whose polytope is simply a square.  The cone over the polytope is shown in figure 3(a), defined by these vectors:
\myfig{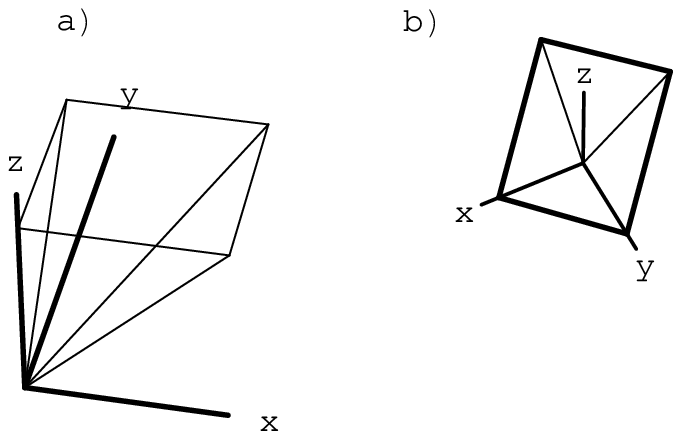}{9}{(a) The toric cone defining the conifold, and the corresponding polytope (simply a unit square) (b) The dual cone for the conifold}
\begin{equation}
v_1=(0,0,1),\quad v_2=(1,0,1),\quad v_3=(1,1,1),\quad v_4=(0,1,1).
\end{equation}
To find the dual cone, we find the inward pointing normal vectors to the sides of the cone, and use those vectors to define the dual lattice cone.  The dual cone for the conifold is shown in figure 3(b), defined by these 4 vectors:
\begin{equation}
\tilde v_1=(0,1,0),\quad \tilde v_2=(-1,0,1),\quad \tilde v_3=(0,-1,1),\quad \tilde v_4=(1,0,0).
\end{equation}
To each of these vectors, we associate a complex coordinate, giving us $\mathbb{C}^4$.  These variables are related multiplicatively based on the linear relationships of the vectors.  In this case, since $\tilde v_1+\tilde v_3=\tilde v_2+\tilde v_4$, then $uv=wz$, where $u$ is associated with $\tilde v_1$, $v$ associated with $\tilde v_3$ etc.  The geometry is defined as the locus of the set of relations among the variables, which in this case is a simple hypersurface since there is only one relation.
\par
In general, steps two and three in this process can be quite difficult.  To illustrate this, we can go ahead and look at the general $Y^{p,q}$ toric diagram, and attempt to calculate the embedding and the number of relations among the variables exactly.
\subsection{Counting relations for the $Y^{p,q}$}
In Appendix B.1 of \cite{BHOP}, the dual cone and generators for the arbitrary $Y^{p,q}$ were calculated.  Restating those results, we found that the dual cone was defined by $4$ vectors:
\begin{equation}
\tilde e_1=(0,0,-1),\quad \tilde e_2=(-p,p,-p+1),\quad\tilde e_3=(-p,-q,q+1),\quad\tilde e_4=(0,-p+q,p-q-1).
\end{equation}
\myfig{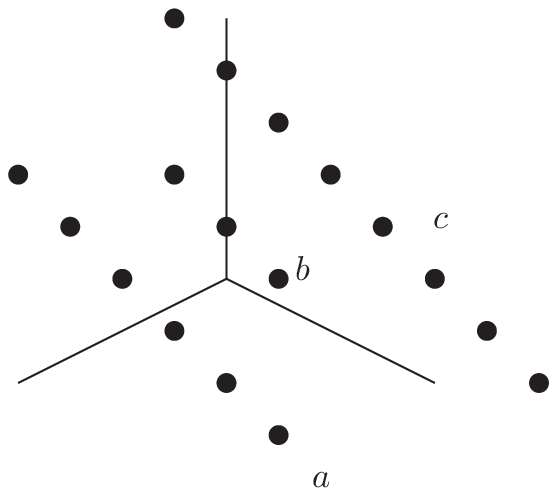}{5}{Sketch of the locations of the generators of the dual cone for $Y^{6,1}$, note that the b generators are beneath the plane corresponding to the lines of a and c generators.}

We found that there were 3 parallel lines of generators. There were $p-q+1$ generators of type $a$ along a line from $\tilde e_1$ to $\tilde e_4$, $3$ generators of type $b$ lying from $(-1,1,-1)$ to $(-1,-1,1)$, and $p+q+1$ generators of type $c$ lying along a line from $\tilde e_2$ to $\tilde e_3$.  This is illustrated for $Y^{6,1}$ in figure 4.
Therefore the geometry can be described as an intersection in $\mathbb{C}^{(2p+5)}$.  To find the specific relations among the $2p+5$ coordinates, however, requires finding all the linear relationships among the generators.  
\par

Restricting ourselves to quadratic relations first, we see that every quadratic relation between variables corresponds to some set of four generators $(a,b,c,d)$ which define the corners of a parallelogram, since that guarantees $a+c=b+d$.  Also note that if the parallelogram is made up of two other parallelograms, the relation is implied by the constituent relations, as illustrated here:
\begin{equation}
\bm{1in}{
\begin{picture}(66,66) (-3,3)
\SetWidth{0.5}
\SetColor{Black}
\Vertex(15,36){2.83}
\Text(-5,26)[lb]{\normalsize{\Black{$a_2$}}}
\Line(0,6)(30,66)
\Line(30,66)(60,66)
\Line(60,66)(30,6)
\Line(0,6)(30,6)
\Vertex(0,6){2.83}
\Text(-20,-4)[lb]{\normalsize{\Black{$a_1$}}}
\Vertex(30,6){2.83}
\Text(10,-6)[lb]{\normalsize{\Black{$b_1$}}}
\Vertex(45,36){2.83}
\Text(25,26)[lb]{\normalsize{\Black{$b_2$}}}
\Vertex(60,66){2.83}
\Text(40,52)[lb]{\normalsize{\Black{$b_3$}}}
\Vertex(30,66){2.83}
\Text(10,56)[lb]{\normalsize{\Black{$a_3$}}}
\end{picture}}=\quad\quad
\bm{1in}{

\begin{picture}(86,66) (-3,3)
\SetWidth{0.5}
\SetColor{Black}
\Vertex(15,36){2.83}
\Text(-5,26)[lb]{\normalsize{\Black{$a_2$}}}
\Line(0,6)(15,36)
\Line(15,36)(60,66)
\Line(60,66)(45,36)
\Line(0,6)(45,36)
\Vertex(0,6){2.83}
\Text(-20,-4)[lb]{\normalsize{\Black{$a_1$}}}
\Vertex(30,6){2.83}
\Text(10,-6)[lb]{\normalsize{\Black{$b_1$}}}
\Vertex(45,36){2.83}
\Text(25,29)[lb]{\normalsize{\Black{$b_2$}}}
\Vertex(60,66){2.83}
\Text(40,60)[lb]{\normalsize{\Black{$b_3$}}}
\Vertex(30,66){2.83}
\Text(10,56)[lb]{\normalsize{\Black{$a_3$}}}
\end{picture}
}+\quad\quad
\bm{1in}{
\begin{picture}(86,66) (-3,3)
\SetWidth{0.5}
\SetColor{Black}
\Vertex(15,36){2.83}
\Text(-5,26)[lb]{\normalsize{\Black{$a_2$}}}
\Line(15,36)(30,66)
\Line(30,66)(45,36)
\Line(45,36)(30,6)
\Line(30,6)(15,36)
\Vertex(0,6){2.83}
\Text(-20,-4)[lb]{\normalsize{\Black{$a_1$}}}
\Vertex(30,6){2.83}
\Text(10,-6)[lb]{\normalsize{\Black{$b_1$}}}
\Vertex(45,36){2.83}
\Text(25,26)[lb]{\normalsize{\Black{$b_2$}}}
\Vertex(60,66){2.83}
\Text(40,52)[lb]{\normalsize{\Black{$b_3$}}}
\Vertex(30,66){2.83}
\Text(10,56)[lb]{\normalsize{\Black{$a_3$}}}
\end{picture}},
\end{equation}
This diagram shows that the relation $a_1b_3=a_3b_1$ is implied by the two relations $a_1b_3=a_2b_2$ and $a_3b_1=a_2b_2$.  Because of this we only have to consider parallelograms where none of the edges intersect a generator.  We refer to these as ``irreducible'' parallelograms.
\par
Now we can proceed to count the total number of quadratic relations:  There are 2 classes of relations, those that only involve one type of generator (only of type $a$, $b$, or $c$), and those that involve relations between two different types.  We will concentrate on the mixed relations first.  Because the lines are parallel, and the generators are equally spaced, this is a simple counting problem.  We need to count the number of ways of choosing two adjacent points in one line, and two adjacent points on the other.  E.g.
\begin{equation}
\bm{1.3in}{
\begin{picture}(96,51) (-3,3)
\SetWidth{0.5}
\SetColor{Black}
\Vertex(15,51){2.83}
\Vertex(45,51){2.83}
\Vertex(75,51){2.83}
\Vertex(30,6){2.83}
\Vertex(60,6){2.83}
\Vertex(0,6){2.83}
\Vertex(90,6){2.83}
\Line(15,51)(0,6)
\Line(15,51)(30,6)
\Line(75,51)(90,6)
\Line(75,51)(60,6)
\Line(75,51)(30,6)
\Line(45,51)(60,6)
\Line(45,51)(0,6)
\Line(45,51)(90,6)
\Line(45,51)(30,6)
\Line(15,51)(60,6)
\Line(15,51)(45,51)
\Line(45,51)(75,51)
\Line(0,6)(90,6)
\end{picture}}\implies(4-1)(3-1)=6.
\end{equation}
Therefore the number of relations between two lines of generators, one of length $m$ and one of length $n$, is simply $(m-1)(n-1)$.  There are $3$ such possible pairings in our case:
\begin{align}
\begin{array}{c|c}
{\rm type}&{\rm \#\; of\; relations}\\
\hline
a/b& 2(p-q)\\\hline
b/c&2(p+q)\\\hline
a/c&p^2-q^2
\end{array}.
\end{align}
The other possibility is relations among the generators of the same type.  This again reduces to a simple counting problem.  All the relations are irreducible (in the above sense), degenerate parallelograms in the line, with vertices at either 3 adjacent points (the center point counted twice), or 4 adjacent points.  Therefore, as long as $m\geq 3$, the number of self relations among a line of $m$ lattice points is simply equal to $(m-2)+(m-3)=2m-5$.  E.g
\begin{equation}
\bm{2in}{
\begin{picture}(138,81) (0,3)
\SetWidth{0.5}
\SetColor{Black}
\Vertex(15,81){2.83}
\Vertex(45,81){2.83}
\Vertex(75,81){2.83}
\Vertex(105,81){2.83}
\Vertex(135,81){2.83}
\Line(15,66)(75,66)
\Line(45,51)(105,51)
\Line(75,36)(135,36)
\Line(15,21)(45,21)
\Line(75,21)(105,21)
\Line(45,6)(75,6)
\Line(105,6)(135,6)
\Vertex(45,66){2.83}
\Vertex(75,51){2.83}
\Vertex(105,36){2.83}
\Vertex(15,66){2.83}
\Vertex(75,66){2.83}
\Vertex(45,51){2.83}
\Vertex(105,51){2.83}
\Vertex(75,36){2.83}
\Vertex(135,36){2.83}
\Vertex(15,21){2.83}
\Vertex(45,21){2.83}
\Vertex(75,21){2.83}
\Vertex(105,21){2.83}
\Vertex(45,6){2.83}
\Vertex(75,6){2.83}
\Vertex(105,6){2.83}
\Vertex(135,6){2.83}
\end{picture}}\implies 2\cdot 5-5=5.
\end{equation}
For our $Y^{p,q}$ case there are 3 such lines.  One with $m=p+q+1$, one with $m=p-q+1$, and one with $m=3$, giving:

\begin{align}
\begin{array}{c|c}
{\rm type}&{\rm \#\; of\; relations}\\\hline
a/a&2p-2q-3\\\hline
b/b&1\\\hline
c/c&2p+2q-3
\end{array}.
\end{align}
So the total number of quadratic relations for the $Y^{p,q}$ is
\begin{align}
\nonumber (2(p-q)+2(p+q)+p^2-q^2)+(2p-2q-3+1+2p+2q-3)=\\
p^2+8p-q^2-5.
\end{align}
This is the total number of quadratic relations (when $p\geq 3$).  However, there are also higher order relations (or extra quadratic relations when $p=2$).
\par
To find the higher order relations, first we note that among a set of generators that lie in a single plane, all relations are quadratic (or reduce to quadratic ones), so we only need to consider relations among all three generator types, $a$, $b$, and $c$.  First we label the various generators explicitly:
\begin{align}
\nonumber a_1=(0,0,-1)\quad&\dots\quad a_{p-q+1}=(0,-p+q,p-q-1)\\
\nonumber b_1=(-1,1,-1)\quad&\dots\quad b_3=(-1,-1,1)\\
c_1=(-p,-q,q+1)\quad&\dots\quad c_{p+q+1}=(-p,p,-p+1).\label{gen}
\end{align} 
By looking at the $x$ coordinates of the $a,b$, and $c$ generators given in (\ref{gen}), we see immediately that the only relations possible relate order $p$ $b$ type generators to a combination of one $a$ and one $c$ type generator.  Also, because of the quadratic relation $b_1b_3=b_2^2$, we can limit ourselves to looking at relations involving $b_1$ or $b_3$, but not both.  Inspecting the generators shows that all relations are of the form
\begin{align}
\nonumber b_1^{p-m}b_2^m&=a_{1+n}c_{p+q+1-l}\\
{\rm for}\quad0\leq m\leq p,\quad n+l=m,\quad0\leq n\leq p-q,&\quad0\leq l\leq p+q
\end{align}
and
\begin{align}
\nonumber b_3^{p-m}b_2^m&=a_{p-q+1-n}c_{1+l}\\
{\rm for}\quad0\leq m\leq p-1,\quad n+l=m,\quad0\leq n\leq p-q,&\quad0\leq l\leq p+q,
\end{align}
Counting the number of these equations is a simple exercise.  We find that
\begin{equation}
\sum_{m=0}^{p-q}(m+1)+q(p-q+1)+\sum_{m=0}^{p-q}(m+1)+(q-1)(p-q+1)=1+2p+p^2-q^2.
\end{equation}
However, each mixed quadratic relation between $a$ and $c$ makes one of these equations redundant, and therefore we must subtract the $p^2-q^2$ number of $a/c$ mixed relations, giving us a total of $1+2p$ order $p$ relations.  Adding this to the number of quadratic relations gives us the total number of relations\footnote{Note that this formula is incorrect for $Y^{1,0}$ (because there are no generators of type $b$ in the conifold), and off by 1 for $Y^{p,p-1}$, or $3$ for $Y^{p,p}$ (because we assumed that $p-q+1\geq 3$.)}
:
\begin{equation}
p^2+10p-q^2-4.
\end{equation}
\par
One thing to note about this formula is its dependence on $p$.  The number of chiral primaries which equals the dimension of the embedding space of the vacuum manifold is simply equal to the number of generators, or $2p+5$.  However the number of relations grows as $p^2-q^2$.  Therefore, in some sense, as $p^2-q^2$ gets large, the intersection becomes more and more constrained, with more and more relations among comparatively fewer number of variables.
\section{Additional Examples}
In this appendix, we will look at two more examples; one trivial and one non-trivial.  We will first derive the deformed conifold using the techniques of Altmann, as a further check of their correctness.  Then we will derive the deformation to the $Y^{3,0}$, which should correspond to a $\mathbb{Z}^3$ orbifold of the conifold.  This example was already studied in \cite{FHU}, however using these new techniques we can find the entire embedding space explicitly. 
\subsection{Deformed conifold}
We already found the generators of the dual cone for the conifold in Appendix A:
\begin{equation}
\tilde v_1=(0,1,0),\quad \tilde v_2=(-1,0,1),\quad \tilde v_3=(0,-1,1),\quad \tilde v_4=(1,0,0).
\end{equation}
Now to describe deformations of this toric variety, we first need to add an extra generator to the cone, corresponding to the vector $\tilde v_5=(0,0,1)$.  If we associate a new variable $t$ to this vector, we see that we now get two relations, $uv=t$ and $wz=t$.  This is a trick which is needed in this special case where $(0,0,1)$ isn't actually a generator of the dual cone.
However, exactly due to the fact that this generator wasn't needed to generate the entire lattice cone, these two equations reduce immediately to the one, original hypersurface equation.  This rewriting enables us to run our algorithm to construct the tautological cone, thus embedding the toric variety in a higher space.  Then we can eliminate $t$ again to determine the final degrees of freedom for the deformations.  
\par
We represent the polytope of the conifold by a sequence of edges:
\begin{equation}
d^1=(1,0),\quad d^2=(0,1),\quad d^3=(-1,0),\quad d^4=(0,-1).
\end{equation}
Using the process outlined in section 4, we associate a point on the original polytope to each of the generators of the dual cone.  We choose 
$(0,0)$ for $\tilde v_1$ and $\tilde v_4$, $(0,1)$ for $\tilde v_3$ and $(1,0)$ for $\tilde v_2$ as the points on the polytope, $a$.  Then we construct paths to each of the four points on the toric polytope, $\lambda^1=\lambda^4=(0,0,0,0)$, $\lambda^2=(1,0,0,0)$, and $\lambda^3=(0,1,0,0)$.  Then finally we construct the weights $\eta$ which end up being the same as the $\lambda$'s.  
Summarizing:
\begin{equation}
\begin{array}{cccc}
{\rm generator\quad}&a&\lambda&\eta\\
\tilde v_1&(0,0)&\quad(0,0,0,0)&\quad(0,0,0,0)\\
\tilde v_2&(1,0)&\quad(1,0,0,0)&\quad(1,0,0,0)\\
\tilde v_3&(0,1)&\quad(0,1,0,0)&\quad(0,1,0,0)\\
\tilde v_4&(0,0)&\quad(0,0,0,0)&\quad(0,0,0,0)
\end{array}.
\end{equation}
Lifting the two relations to the tautological cone gives:
\begin{equation}
uv=t_1,\quad wz=t_2
\end{equation}
However as we found in Appendix A, we have two contraints, $t_1-t_3=0$ and $t_2-t_4=0$ (the quadratic contraints are redundant), so we can parameterize the solutions by 
$t_1=t$ and $t_2=t-s$.  Then 
\begin{equation}
uv=t,\quad wz=t-s.
\end{equation}
Eliminating the extra generator $t$, we simply get $uv=wz+s$, the deformed conifold.  There are many simpler ways of deriving this result, but this method generalizes to varieties which are not complete intersections, while the other, simpler methods rely on the variety being a complete intersection.
\subsection{$Y^{3,0}$, the deformed, $\mathbb{Z}^3$ orbifolded conifold}
For a slightly more complex example, we can look at the $Y^{3,0}$, which we expect to have an unobstructed deformation because it corresponds to a $\mathbb{Z}^3$ orbifold of the conifold.  This is defined as the polytope with vertices at $(0,0)$, $(1,0)$, $(3,3)$, and $(2,3)$.  From our work in Appendix A.1, we know that there are $2p+5=11$ generators of the dual cone, and $p^2+10p-q^2-4=35$ relations among them.  We can list them all explicitly.  First, there are 7 relations that involve only one of the $a$ or $c$ type generators:
\begin{align}
\nonumber a_1a_3=a_2^2,\quad a_2a_4=a_3^2,\quad a_1 a_4=a_2a_3,\quad(a\leftrightarrow c)\\
b_1 b_3=b_2^2.
\end{align}
Then there are 12 relations involving only $a$ and $b$, or $b$ and $c$ type generators:
\begin{align}
\nonumber a_1b_2=a_2b_1,\quad a_2 b_2=a_3 b_1,\quad a_3 b_2= a_4 b_1,\\
a_1 b_3=a_2 b_2,\quad a_2 b_3=a_3 b_2, \quad a_3 b_3=a_4 b_2,\quad(a\leftrightarrow c).
\end{align}
And finally there are 16 cubic relations
\begin{align}
\nonumber b_1^3&=a_1 c_4,\\
\nonumber b_1^2 b_2=a_2 c_4&=a_1 c_3,\\
\nonumber b_1 b_2^2=a_1 c_2=a_2 c_3&=a_3 c_4,\\
\nonumber b_2^3=a_1c_1=a_2c_2=a_3c_3&=a_4c_4,\\
\nonumber b_2^2b_3=a_2 c_1=a_3 c_2&=a_4 c_3,\\
\nonumber b_2b_3^2=a_4 c_2&=a_3 c_1,\\
b_3^3&=a_4c_1.
\end{align}
This gives all the $p^2+10p-q^2-4=35$ relations we expect.  Running the algorithm from section 4, we get this data:
\begin{equation}
\begin{array}{c|c|c|c|c}
{\rm generator}&\eta_0&a()&\lambda&\eta\\\hline
a_1&0&(0,0)&(0,0,0,0)&(0,0,0,0)\\
\hline
a_2&0&(0,0)&(0,0,0,0)&(0,0,0,0)\\
\hline
a_3&0&(0,0)&(0,0,0,0)&(0,0,0,0)\\
\hline
a_4&0&(0,0)&(0,0,0,0)&(0,0,0,0)\\
\hline
b_1&1&(2,3)&(0,1,0,0)&(0,1,0,0)\\
\hline
b_3&1&(1,0)&(1,0,0,0)&(1,0,0,0)\\
\hline
c_1&3&(3,3)&(1,1,0,0)&(3,0,0,0)\\
\hline
c_2&3&(3,3)&(1,1,0,0)&(2,1,0,0)\\
\hline
c_3&3&(3,3)&(1,1,0,0)&(1,2,0,0)\\
\hline
c_4&3&(3,3)&(1,1,0,0)&(0,3,0,0)
\end{array}.
\end{equation}
Then, applying the $\eta$ weights to lift the relations to the tautological cone:
\begin{align}
\nonumber a_1t_2=a_2b_1,\quad a_2 t_2=a_3 b_1,\quad a_3 t_2=a_4 b_1,\\
a_1 b_3=a_2 t_1,\quad a_2b_3=a_3 t_1,\quad a_3b_3=a_4 t_1
\end{align}
is what happens to the $a/b$ relations.
\begin{align}
\nonumber c_1 t_2^2=c_2 b_1 t_1,\quad c_2 t_2^2=c_3 b_1 t_1,\quad c_3 t_2^2=c_4 b_1 t_1,\\
c_1 b_3 t_2=c_2 t_1^2,\quad c_2 b_3 t_2=c_3 t_1^2,\quad c_3 b_3 t_2=c_4 t_1^2
\end{align}
is how the $b/c$ relations are lifted.  Note that they are now cubic.  The $a/a$ and $c/c$ relations are invariant, and the only $b/b$ relation is modified to
\begin{align}
\nonumber a_1a_3=a_2^2,\quad a_2a_4=a_3^2,\quad a_1 a_4=a_2a_3,\quad(a\leftrightarrow c)\\
b_1 b_3=t_1t_2.
\end{align}
And the cubic relations split up like this:
\begin{align}
\nonumber b_1^3&=a_1c_4\\
\nonumber b_1^2t_2=a_2c_4,\quad b_1^2 t_1&=a_1 c_3\\
\nonumber b_1t_1^2=a_1 c_2,\quad b_1 t_1 t_2=a_2 c_3,\quad b_1 t_2^2 &= a_3 c_4\\
\nonumber t_1^3=a_1 c_1,\quad t_1^2 t_2=a_2 c_2,\quad t_1 t_2^2=a_3 c_3,\quad t_2^3&=a_4c_4\\
\nonumber t_1^2b_3=a_2 c_1,\quad t_1t_2 b_3=a_3 c_2,\quad t_2^2 b_3&=a_4c_3\\
\nonumber t_2 b_3^2=a_4 c_2,\quad t_1 b_3^2&=a_3 c_1\\
b_3^3&=a_4 c_1.
\end{align}
Analyzing the ideal simply gives that $t_1$ and $t_2$ are free parameters, identical to the conifold case.  Therefore we relabel $t_1\to b_2$, and $t_2\to (b_2-S)$, giving us our deformed, orbifolded conifold:
\begin{align}
\nonumber a_1(b_2-S)=a_2b_1,\quad a_2 (b_2-S)=a_3 b_1,\quad a_3 (b_2-S)&=a_4 b_1,\\
\nonumber a_1 b_3=a_2 b_2,\quad a_2b_3=a_3 b_2,\quad a_3b_3&=a_4 b_2\\
\nonumber c_1 (b_2-S)^2=c_2 b_1 b_2,\quad c_2 (b_2-S)^2=c_3 b_1 b_2,\quad c_3 (b_2-S)^2&=c_4 b_1 b_2,\\
\nonumber c_1 b_3 (b_2-S)=c_2 b_2^2,\quad c_2 b_3 (b_2-S)=c_3 b_2^2,\quad c_3 b_3 (b_2-S)&=c_4 b_2^2\\
\nonumber a_1a_3=a_2^2,\quad a_2a_4=a_3^2,\quad a_1 a_4=a_2a_3,\quad(a&\leftrightarrow c)\\
\nonumber b_1 b_3&=b_2(b_2-S)\\
\nonumber b_1^3&=a_1c_4\\
\nonumber b_1^2(b_2-S)=a_2c_4,\quad b_1^2 b_2&=a_1 c_3\\
\nonumber b_1b_2^2=a_1 c_2,\quad b_1 b_2 (b_2-S)=a_2 c_3,\quad b_1 (b_2-S)^2 &= a_3 c_4\\
\nonumber b_2^3=a_1 c_1,\quad b_2^2 (b_2-S)=a_2 c_2,\quad b_2 (b_2-S)^2=a_3 c_3,\quad (b_2-S)^3&=a_4c_4\\
\nonumber b_2^2b_3=a_2 c_1,\quad b_2(b_2-S) b_3=a_3 c_2,\quad (b_2-S)^2 b_3&=a_4c_3\\
\nonumber (b_2-S) b_3^2=a_4 c_2,\quad b_s b_3^2&=a_3 c_1\\
b_3^3&=a_4 c_1.
\end{align}
These 35 equations in 11 variables and one parameter $S$ define the geometry of the deformed, $\mathbb{Z}^3$ orbifolded conifold.  The generalization to arbitrary $Y^{p,0}$, though somewhat notationally cumbersome, is straightforward.  This result can be compared with the previous work done in \cite{FHU}.

\end{document}